\documentclass[%
reprint,
superscriptaddress,
 amsmath,amssymb,
 aps,
]{revtex4-1}

\usepackage{graphicx}
\usepackage{dcolumn}
\usepackage{bm}
\usepackage{mathrsfs}

\usepackage[colorlinks,linkcolor=blue,anchorcolor=blue,citecolor=blue,urlcolor=black]%
{hyperref}
\begin{document}

\title{Simulation of topological phases with color center arrays in phononic crystals}
\author{Xiao-Xiao Li}
\affiliation{Shaanxi Province Key Laboratory of Quantum Information and Quantum Optoelectronic Devices,
Department of Applied Physics, Xi'an Jiaotong University, Xi'an 710049, China}
\affiliation{Department of Physics, University of Oregon, Eugene, Oregon 97403, USA}
\author{Bo Li}
\affiliation{Shaanxi Province Key Laboratory of Quantum Information and Quantum Optoelectronic Devices,
Department of Applied Physics, Xi'an Jiaotong University, Xi'an 710049, China}
\author{Peng-Bo Li}
\email{lipengbo@mail.xjtu.edu.cn}
\affiliation{Shaanxi Province Key Laboratory of Quantum Information and Quantum Optoelectronic Devices,
Department of Applied Physics, Xi'an Jiaotong University, Xi'an 710049, China}


\begin{abstract}
We propose an efficient scheme for simulating the topological phases of
matter based on silicon-vacancy (SiV) center arrays in
phononic crystals. This phononic band gap structure allows for long-range spin-spin
interactions with a tunable profile. Under a particular periodic microwave
driving, the band-gap mediated spin-spin interaction can be further designed
with the form of the  Su-Schrieffer-Heeger (SSH) Hamiltonian. In momentum space, we investigate the topological characters of the SSH model, and show that the topological nontrivial phase can be obtained through modulating the periodic driving fields. Furthermore, we explore the zero-energy topological edge states at the boundary of the color center arrays, and study the robust quantum information transfer via the topological edge states. This setup provides a scalable and promising platform for studying topological quantum physics and
quantum information processing with color centers and phononic crystals.
\end{abstract}

\maketitle
\section{introduction}

Topological phases of matter have attracted great interests and developed
potential applications in quantum physics
\cite{nat-496-196,np-10-39,prl-119-023603,sci-359-666,rmp-91-015005,rmp-91-015006,prappl-11-044026,sa-5-7,prl-123-080501}. In particular, topological insulators possess topologically protected surface or edge states, robust to local disorders. The Su-Schrieffer-Heeger (SSH) model, originally derived from the dimerized chain, is the simplest example of a
one-dimensional (1D) topological insulator \cite{prl-42-1698,lecture-2016}.
For the generalized SSH Hamiltonian, the interaction between different
sites of a 1D chain is described by the alternating off-diagonal
elements. Due to the intrinsic topological features of the SSH model,
various quantum systems are employed to simulate the SSH model, and to
explore interesting applications in quantum information processing
\cite{prb-89-085111,Natcomm-6-6710,Natcomm-7-13986,prb-96-125418,prl-119-210401,prl-118-083603,arxiv-n-2019,prappl-12-034014,prb-100-075120,prl-118-076803,prb-97-035442,prl-122-233903,prb-100-075437}%
. However, the simulation of topological phenomena in the quantum domain is
still challenging in practice due to stringent conditions.

The fundamental model of quantum optics is the light-matter interaction
at the single photon level \cite{book-1997}. Photons play a key role in
quantum information science due to their excellent coherence and controllability.
With the advent of quantum acoustics, phonons provide an alternative way to
store and transmit quantum information in hybrid quantum devices
\cite{nsr-2-510}. Moreover, the low speed of phonons enables new dynamic control protocols for
quantum information, and the relatively long acoustic wavelength allows regimes of atomic
physics to be explored that cannot be reached in photonic systems. In
general, there are two types of phonons in quantum systems. One is in the form of the
stationary phonon. In this case, the
vibrational eigenmode of mechanical resonators is taken as the phonon mode
 \cite{SCI-335-1603,Natcomm-6-8603,prappl-4-044003,sr-7-14116,prl-113-020503,Natcomm-5-4429,prl-110-156402}. The other is the propagating phonon, which is resulting from the mechanical
lattice vibration in various acoustic setups, such as surface acoustic wave (SAW) devices
\cite{prb-54-13878,prb-93-041411,prl-116-143602,prx-6-041060,prl-119-180505}, and
phononic crystal structures
 \cite{sci-271-634,prl-86-3012,prl-93-024301,prl-106-084301,sci-343-516,prl-115-104302,opt-3-1404,prl-121-194301}.

Recently, much attention has been paid to the coherent coupling between the
phononic structures and other quantum systems
 \cite{prb-79-041302(R),nat-494-211,sci-346-207,nano-10-55,prl-117-015502,prappl-10-024011,prl-120-213603,prx-8-041027,sci-364-368}. We have investigated the coherent coupling between silicon-vacancy
(SiV) centers and the quantized acoustic modes in a one-dimensional phononic crystal waveguide
 \cite{arxiv-2019}. Color centers in diamond, owing to their long coherence time and excellent optical
properties, have become one of the most promising solid-state quantum emitters \cite{np-7-879,prb-88-064105,prl-121-246402,prl-112-036405,prl-113-263601,prappl-5-044010,prb-97-205444,prl-118-223603}.
The coupling between solid-state spins and quantized acoustic modes in phononic crystals offers new
paradigm for investigating spin-phonon interactions near the phonon band-gap.
In analogy to photonic crystals, phononic crystals are constructed with elastic waves propagating in periodic
structures modulated by periodic elastic modula and mass densities. For
phononic crystals, one of the most prominent features is the presence of
band-gap structures, which provide stronger interactions due to the much
tighter confinement of the mediating phonon
 \cite{prb-49-2313,nat-462-78}. More importantly,
phononic crystals offer a promising platform for practical quantum technologies
because of the extremely low thermoelastic mechanical dissipation.

In this work, we present a periodic driving protocol to simulate topological
phases with a color center-phononic crystal system in both the one-dimensional (1D) and 2D cases. In the setup,  SiV center arrays are
coupled to the quantized modes of a  phononic crystal near the band-gap. With the band gap
engineered spin-phonon interaction, the phononic crystal modes are
distributed around the spins with an exponentially decaying envelope. We
show that the SSH-type Hamiltonian can be obtained by applying periodic
microwave driving fields to the SiV spins. Then we explore the topological
properties of the effective spin-spin system in the momentum space.
Furthermore, we also study the zero-energy topological edge states at the boundary of the color center array, and show the robust quantum information transfer via the topological edge states. Compared with other nanomechanical systems,
phononic crystals possess unique band gap structures and exceptional physical
properties, providing an ideal interface with diamond defect spins.
Moreover, suitably designed periodic driving fields enable the highly controllable and tunable
SSH model in SiV center arrays. Our results allow to further explore the
topological properties of the spin-phonon systems, and also open up new ways
for quantum information processing in phononic crystal systems.

\section{Simulation of the 1D SSH model }
\subsection{The setup}

\begin{figure}[tbp]
\includegraphics[width=8.6cm]{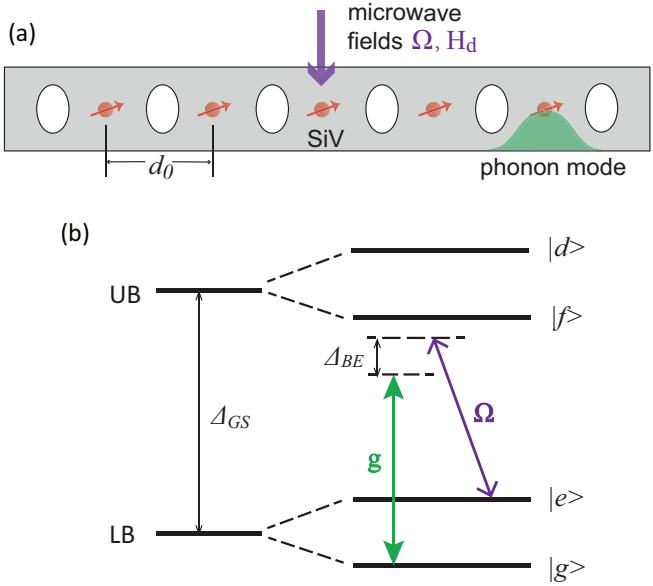}
\caption{(Color online) (a) Schematics of the hybrid device studied in this
work. An array of SiV centers are implanted evenly in a 1D
phononic crystal waveguide. $d_{0}$ is the distance between the adjacent
spins. The spins are driven by microwave field $\Omega$ and the periodic driving fields $\hat{H}_{d}$.
(b) Simplified energy levels of a single SiV center in the electronic ground
state. Here, $g$ describes the spin-phonon interaction, $\Delta _{BE}$ is
the detuning between the effective spin transition and the phononic band
edge frequency.}
\label{fig_model}
\end{figure}

We first consider a spin-phononic crystal system as shown in Fig.~$1(a)$, an array of $2N$ SiV
centers are implanted evenly in a 1D phononic crystal
waveguide. The generalization of this 1D scheme to the case of 2D will be considered later. The diamond waveguide is perforated with periodic elliptical air
holes, which provide the tunable phononic band structure. In general, the
phononic crystal supports acoustic guide modes $\omega _{n,k}$, where $n$ is
the band index and $k$ is the wave vector along the waveguide direction. The
mechanical displacement mode profile $\vec{Q}(\vec{r},t)$ can be obtained by
solving the elastic wave equation \cite{TE-1986}. Analogous to the
electromagnetic field in quantum optics, the mechanical displacement field
can be quantized, i.e., $H_{p}=\sum_{n,k}\hbar \omega _{n,k}a%
_{n,k}^{\dagger }a_{n,k}$, with $a_{n,k}$ and $a%
_{n,k}^{\dagger }$ the annihilation and creation operators for the phonon
modes.
\begin{figure}[tbp]
\includegraphics[width=8.6cm]{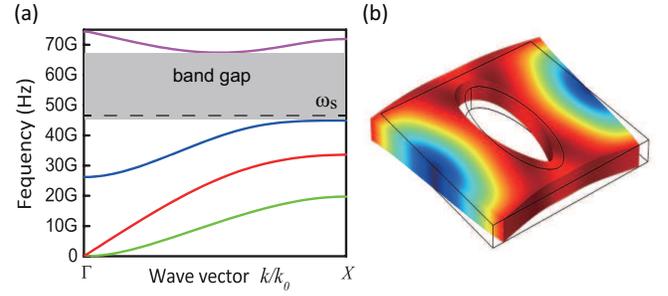}
\caption{(Color online) (a) The phononic dispersion relation for the four
lowest bands. The gray area denotes the phonon bandgap, and $\protect\omega _{s}$
is the effective spin transition frequency. (b) The displacement pattern
corresponding to the third band.}
\label{fig_energy}
\end{figure}

The coherent interaction between phononic crystal modes and electron spin states
of a SiV center has been studied in our previous work \cite{arxiv-2019}. For the
SiV center in diamond, the electronic ground state is split by the
spin-orbit interaction and crystal strain into a lower branch (LB) and upper
branch (UB) separated by $\Delta _{GS}=46$ GHz. In the presence of an
external magnetic field, each branch is further broken to reveal two
sublevels, i.e., $\{\left\vert g\right\rangle =\left\vert e_{-}\downarrow
\right\rangle ,\left\vert e\right\rangle =\left\vert e_{+}\uparrow
\right\rangle \}$ and $\{\left\vert f\right\rangle =\left\vert
e_{+}\downarrow \right\rangle ,\left\vert d\right\rangle =\left\vert
e_{-}\uparrow \right\rangle \}$, as show in Fig.~$1(b)$, where $\left\vert
e_{\pm }\right\rangle $ are eigenstates of the orbital angular momentum
operator. When the transition frequency of the spin state is tunned close to
the phononic band edge, we obtain the strong strain coupling between the SiV
center and the phononic crystal mode. Applying a microwave driving field $%
\Omega $ to couple the states $\left\vert e\right\rangle $ and $\left\vert
f\right\rangle $ and define appropriate detunings, the spin-phonon
Hamiltonian can be mapped to the Jaynes-Cummings model,
namely
\begin{equation}
H_{s-p}=\sum_{k}\hbar \omega _{k}a_{k}^{\dagger }a%
_{k}+\hbar \omega _{s}\sigma_{ee}+\underset{k}{\sum }\hbar g(a%
_{k}\sigma_{eg}e^{ikx_{0}}+H.c.),
\end{equation}%
where $\sigma_{eg}=\left\vert e\right\rangle \left\langle g\right\vert
$, $\omega _{s}$ is the effective spin transition frequency, and $g$ is the
effective spin-phonon coupling strength. Here, we assume that the defect
center is only coupled to a single band of the phononic crystal, so
the index $n$ can be omitted. In Figure.~$2$, we numerically simulated the
mechanical band structure and displacement pattern of one unit cell by finite
element method (FEM), which is performed with the COMSOL Multiphysics
software. By designing the parameters of the phononic crystal, the spin
transition frequency is exactly lies within a phononic bandgap.

For a single excitation in the system, there exists a bound state $|\psi
_{b}\rangle =\cos \theta |0\rangle |e\rangle +\sin \theta |1\rangle
|g\rangle $ within the phononic bandgap. Here $|0\rangle $ is the vacuum state of
the phonon mode, and $|1\rangle =\int dkc_{k}a_{k}^{\dagger }|0\rangle
$ is single excitation state for the phonon modes. The bound state satisfies
the eigenvalue equation $H_{s-p}|\psi _{b}\rangle =\hbar \omega
_{b}|\psi _{b}\rangle $, where $\omega _{b}$ is the corresponding eigenfrequency.
Based on the eigenvalue equation, the phononic spatial mode has the
following form
\begin{equation}
\varepsilon (x)=\int dkc_{k}Q_{k}(x)=\sqrt{\frac{2\pi }{L_{c}}}%
e^{-|x-x_{0}|/L_{c}}Q_{k_{0}}(x),
\end{equation}%
that is the phononic part of the bound state is exponentially localized around
the spin, with $L_{c}$ the localized length of the phononic wavefunction.

In the following, we study the interaction between phononic crystal waveguide modes
and an array of SiV spins. Here we assume that the SiV centers are equally coupled to
the phononic mode near the band gap, and the direct spin-spin interaction can
be neglected, since it is excessively week compared with the spin-phonon
interaction. Thus the interaction Hamiltonian between the defect spins and
phonon modes can be expressed as
\begin{equation}
H_{s-s}=\sum_{j,k}\hbar g(a_{k}%
\sigma_{eg}^{j}e^{i\delta _{k}t+ikx_{j}}+H.c.),
\end{equation}%
with $\delta
_{k}=\omega _{s}-\omega _{k}$. Assuming the large detuning regime, $\delta
_{k}\gg g$, we can adiabatically eliminate the phonon modes \cite{cjp-85-625}%
, and then get the effective Hamiltonian
\begin{equation}
H_{s-s}=\overset{2N}{\sum_{i,j=1}}\hbar \mathcal{M}_{i,j}(\sigma_{eg}^{i}%
\sigma_{ge}^{j}+H.c.),
\end{equation}%
where
\begin{equation}
\mathcal{M}_{i,j}=\frac{g_{c}^{2}}{2\Delta _{BE}}e^{-|x_{i}-x_{j}|/L_{c}}
\end{equation}%
denotes the effective spin-spin interaction, $g_{c}=g\sqrt{2\pi a/L_{c}}$ is
the the band gap
engineered spin-phonon coupling strength, and $\Delta _{BE}=\omega _{s}-\omega _{BE}$,
with $\omega _{BE}$ the phononic band edge frequency. Note that different
from the conventional dipole-dipole interaction mediated by a mechanical resonator or waveguide, the band-gap mediated
spin-spin interaction is decay exponentially with the distance between spins. The detailed derivation can be found in Ref. \cite{arxiv-2019}.

\subsection{The Periodic driving}

The periodic driving is known to render effective Hamiltonian in which
specific terms can be adiabatically eliminated. Driving a quantum system periodically in time can profoundly alter its long-time dynamics and trigger topological order \cite{prx-4-031027}. As for the SiV color center in diamond, the electric structure is comprised of spin and orbital degrees of freedom. Considering the spin-orbit interaction and strain environment, there are four sublevels combined by orbital and spin components, as shown in Fig.~$1(b)$. The spin-flip transitions are allowed between ground-state levels of opposite electronic
spin in the SiV center. Thus, we can define the two
lower sublevels $(|g\rangle ,|e\rangle )$ as a spin qubit, and
apply the extra driving
fields to the SiV centers \cite{prl-119-210401}
\begin{equation}
H_{d}=\frac{\hbar \Omega _{d}}{2}\overset{2N}{\sum_{j=1}}\sigma _{j}^{z}+%
\frac{\hbar \eta \omega _{d}}{2}\cos (\omega _{d}t)\overset{2N}{\sum_{j=1}}%
\cos (\Delta kx_{j}+\omega _{d}t)\sigma _{j}^{z},
\end{equation}%
with the Pauli operator component $\sigma _{j}^{z}=|e\rangle _{j}\langle
e|-|g\rangle _{j}\langle g|$. The first term describes a transverse
microwave field. The second term represents a periodically driving, which can
be realized by a time-dependent standing wave with frequency $\omega _{d}$
and wavevector $\Delta k$ along the array direction. $\eta $ is the
dimensionless coupling strength, $x_{j}=d_{0}j$ is the equilibrium position
of the spin, with $d_{0}$ the distance between the evenly spaced spins.

Now we transform the total Hamiltonian $H_{tot}=H_{s-s}+H_{d}$ into
the interaction picture, with the unitary operator $%
U(t)=e^{-i\int_{0}^{t}d\tau H_{d}/\hbar }$. Here we note that $H_{d}$ is
time-dependent fields. In the interaction picture,
\begin{eqnarray}
\sigma _{eg}^{j}&\rightarrow& e^{i\Delta _{j}(t)\sigma _{j}^{z}}\sigma _{eg}^{j}e^{-i\Delta _{j}(t)\sigma _{j}^{z}}=\sigma _{eg}^{j}e^{2i\Delta _{j}(t)}, \notag\\
\sigma _{ge}^{j}&\rightarrow& e^{i\Delta _{j}(t)\sigma _{j}^{z}}\sigma _{ge}^{j}e^{-i\Delta _{j}(t)\sigma _{j}^{z}}=\sigma _{ge}^{j}e^{-2i\Delta _{j}(t)},
\end{eqnarray}%
with%
\begin{equation}
\Delta _{j}(t)=\frac{\Omega _{d}}{2}t+\frac{\eta \omega _{d}}{2}\cos (\Delta kx_{j}+\omega _{d}t)\sin (\omega _{d}t).
\end{equation}%
To simplify the results, we use the trigonometric identity $\cos A-\cos
B=-2\sin (\frac{A+B}{2})\sin (\frac{A-B}{2})$ and Jacobi-Anger expansion $%
e^{iz\sin \phi }=\sum_{n=-\infty }^{\infty }B_{n}(z)e^{in\phi }$, where $%
B_{n}(z)$ is the Bessel functions of the first kind. Here we consider the
limit $\omega _{d}\gg \mathcal{M}_{i,j}$. Under the rotating wave approximation, only
terms that contain zero-order Bessel functions are remained, ie., $n=0$.
Then the interaction Hamiltonian has the form %
\begin{equation}
H_{tot}=\overset{2N}{\sum_{i,j=1}}\hbar\mathcal{M}_{i,j}J_{i,j}(\sigma _{eg}^{i}\sigma
_{ge}^{j}+\sigma _{ge}^{i}\sigma _{eg}^{j}),
\end{equation}%
with%
\begin{equation}
J_{i,j}=B_{0}(2\eta \sin (\frac{\pi }{4}(i+j)+\omega _{d}t)\sin \frac{\pi }{4}%
(i-j)).
\end{equation}%
To achieve the periodic coupling, here we fixed $\Delta k=\frac{\pi }{2d_{0}}
$. According to the expression of Bessel function, we can get $%
J_{j,j+1}=J_{j+2,j+3}$. Furthermore, for the nearest-neighbor spins, the
interaction strength $\mathcal{M}=\frac{g_{c}^{2}}{2\Delta _{BE}}%
e^{-d_{0}/L_{c}}$ with $d_{0}=\left\vert x_{i}-x_{i+1}\right\vert $. If we
define
\begin{equation}
\delta =\frac{J_{2,3}-J_{1,2}}{J_{2,3}+J_{1,2}},
\end{equation}%
then the Hamiltonian $H_{tot}$ can be rewritten as%
\begin{equation}
H_{tot}=\overset{2N}{\sum_{j=1}}\hbar J\sigma _{eg}^{j}\sigma _{ge}^{j+1}+H.c.,
\end{equation}%
with $J=\mathcal{M}[1+(-1)^{j}\delta ]$. As sketched in Fig.~$3(a)$, two possible
coupling rates $J_{1}=\mathcal{M}(1-\delta )$ and $J_{2}=\mathcal{M}(1+\delta )$ are staggered
along the array. To better describe the physical picture of Eq.~$(12)$, we
rewrite the coupling pattern of the Fig.~$3(a)$ as%
\begin{equation}
H_{tot}=\overset{2N}{\underset{j\in odd}{\sum }}\hbar J_{1}\sigma _{eg}^{j}\sigma
_{ge}^{j+1}+\overset{2N}{\underset{j\in even}{\sum }}\hbar J_{2}\sigma
_{eg}^{j}\sigma _{ge}^{j+1}+H.c..
\end{equation}

Considering this periodic spin-spin interaction, we group the nearest-neighbor spins with the coupling strength $J_{1}$ into a unit cell, where odd spins are labeled as $A_{n}$ and even spins are labeled as $B_{n}$, $n=1,2,\ldots,N$. The interaction Hamiltonian becomes
\begin{equation}
H_{SSH}=\overset{N}{\underset{n=1}{\sum }}\hbar(J_{1}A_{n}^{\dagger
}B_{n}+J_{2}B_{n}^{\dagger }A_{n+1}+H.c.),
\end{equation}%
with%
\begin{eqnarray}
A_{n} &=&\sigma _{ge}^{j},\ j=1,3,5,7\ldots,  \notag \\
B_{n} &=&\sigma _{ge}^{j},\ j=2,4,6,8\ldots.
\end{eqnarray}%
This is the well-known one-dimensional SSH model, where $J_{1}$ and $J_{2}$ describe the intracell
and intercell hoppings, respectively. The SSH model occurs naturally in many
solid-state systems, e.g., polyacetylene, which is known as the simplest
instances of a topological insulator. Likewise, the staggering of the
hopping amplitudes has also been realized in several other quantum systems, such
as optical cavities, trapped-ions and superconducting circuits \cite{prl-119-210401,prl-118-083603}. In our work, phononic crystals possess unique band gap
structures, which provide stronger spin-phonon interactions due to the much
tighter confinement of the mediating phonon. Moreover, periodic driving fields enable the
highly controllable SSH model in the color center arrays.

\subsection{Topological characters}

\begin{figure}[tbp]
\includegraphics[width=8.6cm]{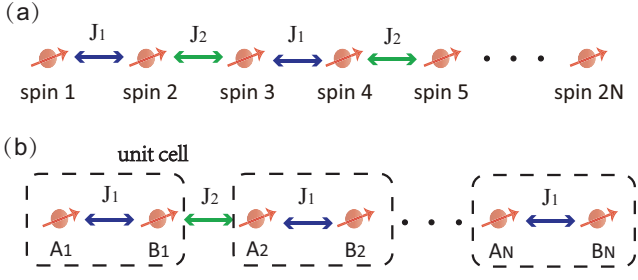}
\caption{(Color online) (a) Schematic diagram for the array of coupled SiV
spins with staggering coupling strength $J_{1}$ and $J_{2}$. (b) The
schematic diagram of Hamiltonian $H_{SSH}$, $J_{1}$ and $J_{2}$ describe the
intracell and intercell hoppings, respectively.}
\label{fig_effective}
\end{figure}

The SSH model has served as a prototypical example of the one-dimensional
system supporting topological character. To explore topological features of
the effective spin-spin system, we convert $H_{tot}$ to the momentum space.
Considering periodic boundary conditions, we can make the Fourier
transformation
\begin{figure}[tbp]
\includegraphics[width=8.6cm]{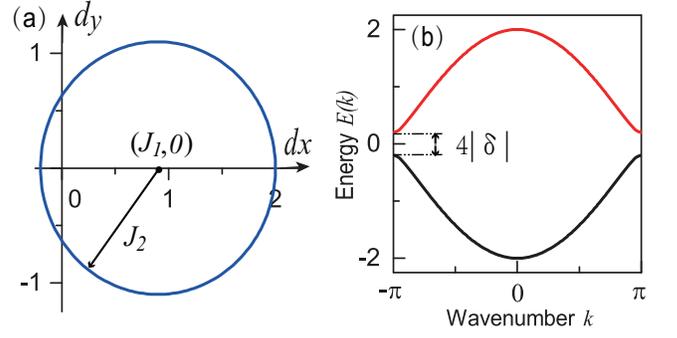}
\caption{(Color online) (a) The path of the endpoints of the vector $d_{k}$
for $\protect\delta =0.1$. (b) The corresponding dispersion relations of the
SSH model in momentum space. The other parameter is $\mathcal{M}=1$. }
\label{fig_energy}
\end{figure}
\begin{eqnarray}
A_{n} &=&\frac{1}{\sqrt{N}}\sum_{k}e^{ink}A_{k},  \notag \\
B_{n} &=&\frac{1}{\sqrt{N}}\sum_{k}e^{ink}B_{k},
\end{eqnarray}%
where $k=2\pi m/N(m=1,2,...,N)$ is the wavenumber in the first Brillouin
zone, and $A_{k}$ and $B_{k}$ are the momentum space operators. Defining the
unitary operator $\psi _{k}=\left(
\begin{array}{cc}
A_{k} & B_{k}%
\end{array}%
\right) ^{T}$, the Hamiltonian $H_{SSH}$ can be rewritten as%
\begin{equation}
H_{SSH}=\sum_{k}\psi _{k}^{\dagger }H(k)\psi _{k},
\end{equation}%
where
\begin{equation}
H(k)=\hbar\left(
\begin{array}{cc}
0 & f(k) \\
f^{\ast }(k) & 0%
\end{array}%
\right)
\end{equation}%
is the momentum-space Hamiltonian. Here, $f(k)=d(k)\cdot \sigma$
describes the coupling between $A$ and $B$ spins in momentum space, $%
\sigma=(\sigma _{x},\sigma _{y},\sigma _{z})$ is the Pauli matrix, and $%
d(k)$ denotes a three-dimensional vector field. For the generalized
SSH model, we have
\begin{eqnarray}
d_{x}(k) &=&J_{1}+J_{2}\cos k,  \notag \\
d_{y}(k) &=&J_{2}\sin k,  \notag \\
d_{z}(k) &=&0.
\end{eqnarray}%
We show the path of the endpoints of the vector $d(k)$ for different $%
\delta $ in Fig.~$4(a)$. As the wavenumber runs through the Brillouin zone $%
k=0\rightarrow 2\pi $, the path depicted by the endpoint of $d(k)$ is
a closed circle of radius $J_{2}$ on the $dx$-$dy$ plane, centered at $%
(J_{1},0)$.

Now we proceed to investigate the energy spectrum of the SSH model in
momentum space. Sloving the eigenvalue equation%
\begin{equation}
H(k)\psi _{k}=E(k)\psi _{k},
\end{equation}%
we get
\begin{equation}
E(k)=\pm \hbar \left\vert J_{1}+J_{2}e^{-ik}\right\vert .
\end{equation}%
Furthermore, the eigenenergy can be expressed as $E(k)=\pm \hbar \mathcal{M}\sqrt{2(1+\delta
^{2})+2(1-\delta ^{2})\cos k}$. Figs.~$4(b)$ shows the
corresponding dispersion relations for different $\delta $: (i) for the general
case with $\delta \neq 0$, the spin-spin interactions have a chiral symmetry, such that all eigenmodes can be grouped in chiral symmetric pair with opposite energies. Thus, the engrgy spectrum is split into two branches,
and there exist a band gap $2\Delta E(k)$ locates at $k=\pi $, with%
\begin{equation}
\Delta E(k)=\min E(k)=2\left\vert \delta \right\vert .
\end{equation}%
(ii) for the case with $\delta =0$, i.e., the spin-spin hopping rate is a
constant, the band gap is closed, which recovers the normal 1D tight-binding
model.

From Eq.~$(21)$ we conclude that, the system with $\delta <0$ and $%
\delta >0$ share the identical band structure, but they are topologically inequivalent.
As a topological invariant, the Winding number can be used to characterize
the topological properties of the one-dimensional system. According to the
above derivation, the energy bands of the Hamiltonian $E(k)$ is determined
by the vector field $d(k)$. Hence, the Winding number can be
conveniently expressed as%
\begin{equation}
\mathcal{W}=\frac{1}{2\pi }\int_{0}^{2\pi }(n_{x}\partial _{k}n_{y}-n_{y}\partial
_{k}n_{x})dk,
\end{equation}%
where $(n_{x},n_{y})=(d_{x},d_{y})/\sqrt{d_{x}^{2}+d_{y}^{2}}$ is the
normalized vector, and $\partial _{k}$ is the partial derivative with
respect to $k$. So we can obtain that the Winding number is either $0$ or $1$%
. In the case $\delta <0$, the Winding number $\mathcal{W}=0$, the system is topological trivial. While in the case $%
\delta >0$, the Winding number $\mathcal{W}=1$, the system has a topological nontrivial phase.Thus, at the critical point $\delta =0$, one can implement the topological phase transitions.

\section{Simulation of the 2D SSH model}

The Su-Schrieffer-Heeger model inherently possesses topological features,
providing an effective pattern to study topological phenomena in quantum
systems. It is naturally expected to investigate the SSH model in
high-dimensional quantum systems and develop interesting applications in
quantum information processing.
For phononic crystals, owing to the advantage of the scalable nature of nanofabrication, the extension to
high-dimensional cases is experimentally feasible and has been extensively studied. In this section, we focus on the 2D SSH model and relevant
topological characters in this spin-phononic system.

\subsection{The setup}
\begin{figure}[tbp]
\includegraphics[width=8cm]{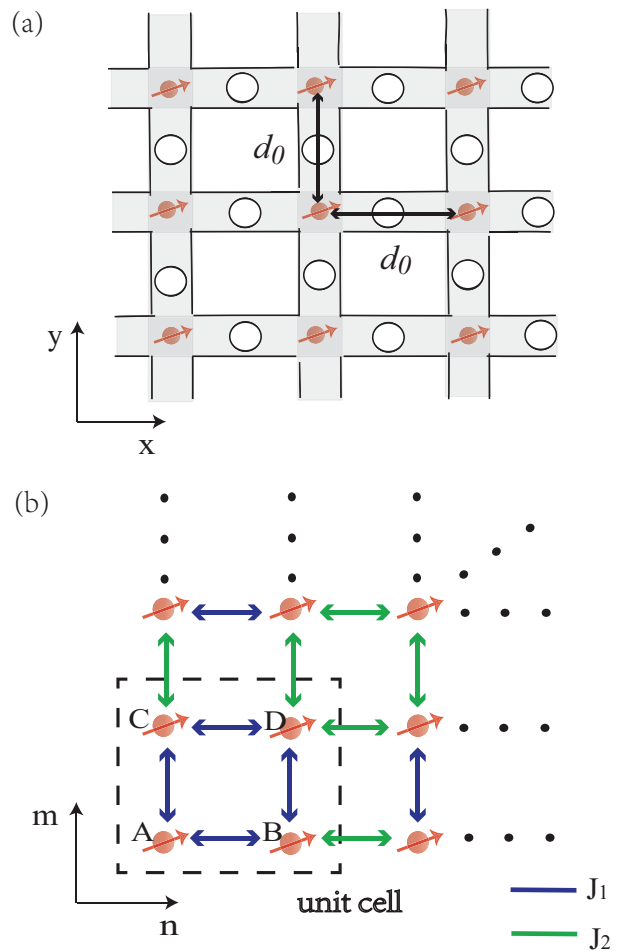}
\caption{(Color online) (a) Schematics of the two-dimensional phononic network studied in this
work. SiV
spins are located at the nodes of the phononic structure. $d_{0}$ is the
distance between two adjacent spins. (b) The
schematic diagram of the Hamiltonian $H_{SSH}^{(2D)}$. There are four spins in each unit cell,
labeled as \{$A,B,C,D$\}, with $J_{1}$ and $J_{2}$ described the
intracell and intercell hoppings, respectively.}
\label{fig_energy}
\end{figure}

Here we consider a
phononic network with square lattices on the $x$-$y$ plane, with $2N\times 2N
$ SiV spins located separately at the nodes of the phononic structure, as
depicted in Fig.~$5(a)$. Based on the coupling of SiV center arrays to the
1D phononic crystal, we thereby obtain the phononic mediated spin-spin
interactions in this 2D phononic network
\begin{eqnarray}
H_{s-s}^{(2D)} &=&H_{s-s}^{(x)}+H_{s-s}^{(y)}, \notag \\
H_{s-s}^{(x)}&=&\overset{2N}{\sum_{l=1}}\overset{2N}{\sum_{i,j=1}}\hbar(\mathcal{M}%
_{i,j}\sigma_{eg}^{(i,l)}\sigma_{ge}^{(j,l)}+H.c.), \notag \\
H_{s-s}^{(y)}&=&\overset{2N}{\sum_{j=1}}\overset{2N}{\sum_{k,l=1}}\hbar(\mathcal{M}_{k,l}%
\sigma_{eg}^{(j,k)}\sigma_{ge}^{(j,l)}+H.c.),
\end{eqnarray}%
where $H_{s-s}^{(x)}$ and $H_{s-s}^{(y)}$ describe the effective spin-spin
interactions in the $x$ and $y$ directions, respectively, and $\mathcal{M}_{i,j},%
\mathcal{M}_{k,l}$ are the corresponding phonon mediated spin-spin hopping
rates.

\subsection{The periodic driving}

According to the 1D SSH model, we can obtain a topological nontrivial system
by applying a specific periodic driving to the SiV spins. For the 2D case,
we consider adding two mutually perpendicular microwave fields to the color
center arrays \cite{prb-95-205125}. The first one is a time-dependent microwave field of frequency $\omega _{d}$ in the $x$
direction. The other is an identical periodic driving in the $y$ direction.
These two periodic driving terms have the form
\begin{eqnarray}
H_{d}^{(x)} &=&\frac{\hbar \eta \omega _{d}}{2}\cos (\omega _{d}t)\overset{2N}{\sum_{l=1}}\overset{2N%
}{\sum_{j=1}}\cos (\Delta k_{x}x_{j}+\omega _{d}t)\sigma _{j,l}^{z},  \notag
\\
H_{d}^{(y)} &=&\frac{\hbar \eta \omega _{d}}{2}\cos (\omega _{d}t)\overset{2N}{\sum_{j=1}}\overset{2N%
}{\sum_{l=1}}\cos (\Delta k_{y}y_{l}+\omega _{d}t)\sigma _{j,l}^{z},
\end{eqnarray}
where $\Delta {k}_{x}$ and $\Delta {k}_{y}$ are the wavevectors along the spin array
direction. $(x_{j},y_{l})$ describes the spin position in the phononic
network, here $x_{j}=d_{0}j$ and $y_{l}=d_{0}l$. In addition, for the SiV
spins, we apply an additional transverse microwave driving $H_{d}^{0}=\frac{%
\hbar \Omega _{d}}{2}\sum_{j,l}\sigma _{j,l}^{z}$, which enables the external
driving of spins to have the same form as Eq.~$(6)$.

For the periodic driving spin arrays along the $x$ direction, the total Hamiltonian is
given by%
\begin{equation}
H_{tot}^{(x)}=H_{s-s}^{(x)}+H_{d}^{(x)}+H_{d}^{0}.
\end{equation}%
In the interaction picture, we introduce the unitary operator $%
U(t)=e^{-i\int_{0}^{t}d\tau (H_{d}^{(x)}+H_{d}^{0})/\hbar }$. After the
unitary transformation, we can obtain%
\begin{eqnarray}
\sigma _{eg}^{(j,l)} &\rightarrow &\sigma _{eg}^{(j,l)}e^{2i\Delta _{j}(t)},
\notag \\
\sigma _{ge}^{(j,l)} &\rightarrow &\sigma _{ge}^{(j,l)}e^{-2i\Delta _{j}(t)},
\end{eqnarray}%
with%
\begin{equation}
\Delta _{j}(t)=\frac{\Omega _{d}}{2}t+\frac{\eta \omega _{d}}{2}\cos (\Delta
k_{x}x_{j}+\omega _{d}t)\sin (\omega _{d}t).
\end{equation}%
In the regime $\omega _{d}\gg \mathcal{M}_{i,j}$, we proceed to
renormalize the spin-spin hopping amplitudes with the zero-order Bessel
function, and obtain the interaction Hamiltonian%
\begin{equation}
H_{tot}^{(x)}=\overset{2N}{\sum_{l=1}}\overset{2N}{\sum_{i,j=1}}\hbar \mathcal{M}_{i,j}J_{i,j}(\sigma
_{eg}^{i,l}\sigma _{ge}^{j,l}+\sigma _{ge}^{i,l}\sigma _{eg}^{j,l}),
\end{equation}%
with
\begin{equation}
J_{i,j}=B_{0}(2\eta \sin (\frac{\pi }{4}(i+j)+\omega _{d}t)\sin \frac{\pi }{4%
}(i-j)).
\end{equation}%
Similar to the discussion in the $x$ direction, the Hamiltonian in the $y$
direction can be written as%
\begin{equation}
H_{tot}^{(y)}=H_{s-s}^{(y)}+H_{d}^{(y)}+H_{d}^{0}.
\end{equation}%
In the regime $\omega _{d}\gg \mathcal{M}_{k,l}$, we obtain the effective
spin-spin interaction along the $y$ coordinate
\begin{equation}
H_{tot}^{(y)}=\overset{2N}{\sum_{j=1}}\overset{2N}{\sum_{k,l=1}}\hbar \mathcal{M}_{k,l}J_{k,l}(\sigma
_{eg}^{(j,k)}\sigma _{ge}^{(j,l)}+\sigma _{ge}^{(j,k)}\sigma _{eg}^{(j,l)}),
\end{equation}%
with%
\begin{equation}
J_{k,l}=B_{0}(2\eta \sin (\frac{\pi }{4}(k+l)+\omega _{d}t)\sin \frac{\pi }{4%
}(k-l)).
\end{equation}%
To simplify the model, here we assumed $\Delta {k}_{x}=\Delta {k}_{y}=\frac{%
\pi }{2d_{0}}$. In this case, according to the definition of $\delta $ given
in Eq.~$(11)$, we have
\begin{equation}
\delta _{x}=\delta _{y}=\delta.
\end{equation}%
For the 2D spin-spin interactions, the total Hamiltonian can be written as%
\begin{equation}
H_{tot}^{(2D)}=\overset{2N}{\sum_{j=1}}\overset{2N}{\sum_{l=1}}\hbar(J_{x}\sigma
_{eg}^{(j,l)}\sigma _{ge}^{(j+1,l)}+J_{y}\sigma _{eg}^{(j,l)}\sigma
_{ge}^{(j,l+1)}+H.c.),
\end{equation}%
with $J_{x}=\mathcal{M}[1+(-1)^{j}\delta ]$ and $J_{y}=\mathcal{M}[1+(-1)^{l}\delta ]$. Therefore, there
are two possible coupling constants $J_{1}$ and $J_{2}$  staggered in the
$x$ and $y$ directions.

In the following, analogous to Eq.~$(14)$, we group the
nearest-neighbor spins with the coupling strength $J_{1}$ into a unit cell.
Then we get a two-dimensional system with $N\times N$ unit cells.
As shown in Fig.~$5(b)$, there are four spins in a unit cell, which are
labeled as \{$A,B,C,D$\}, respectively. Then we can rewrite the two-dimensional
Hamiltonian as%
\begin{eqnarray}
H_{SSH}^{(2D)} &=&\overset{N}{\sum_{n=1}}\overset{N}{\sum_{m=1}}%
\hbar[J_{1}(A_{n,m}^{\dagger }B_{n,m}+A_{n,m}^{\dagger }C_{n,m} \notag \\
&&+B_{n,m}^{\dagger }D_{n,m}+C_{n,m}^{\dagger }D_{n,m}+H.c.) \notag \\
&&+J_{2}(A_{n+1,m}^{\dagger }B_{n,m}+A_{n,m+1}^{\dagger }C_{n,m} \notag \\
&&+B_{n,m+1}^{\dagger }D_{n,m}+C_{n+1,m}^{\dagger }D_{n,m}+H.c.)].
\end{eqnarray}
This is the generalized two-dimensional SSH model \cite{prappl-12-034014,prb-100-075120,prl-118-076803}. For simplicity, here we
introduce $(n,m)$ to describe the position of each unit cell, $n,m=1,2,\ldots ,N$.

\subsection{Topological characters}

\begin{figure}[tbp]
\includegraphics[width=8.6cm]{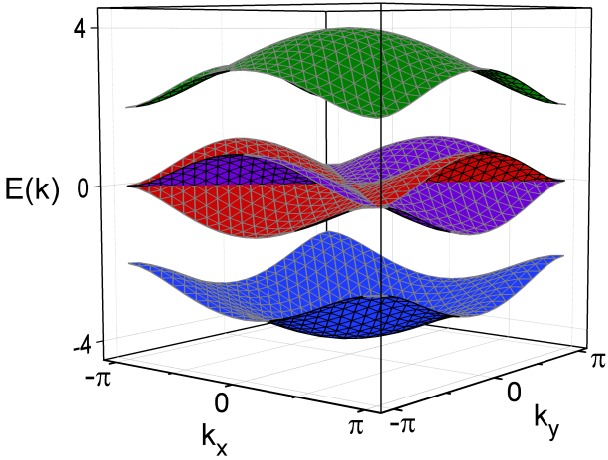}
\caption{(Color online) Band structure of the 2D SSH model as a function of $k_{x}
$ and $k_{y}$,
with $\protect\delta =0.5$. The energy spectrum has four branches.}
\label{fig_energy}
\end{figure}

To explore the topological features of the two-dimensional SSH model, we proceed
to convert $H_{tot}$ to the momentum space. As displayed in Fig.~5, four nearest-neighbor SiV spins with the coupling strength $J_{1}$ form a unit cell, which is a
square lattice geometry.  The
distance between two adjacent spins is $d_{0}$. Thus, the primitive
translation vectors are $\textbf{a$_{1}$}=(2d_{0},0)$ and $\textbf{a$_{2}$}=(0,2d_{0})$, and the
corresponding reciprocal lattice vectors are $\textbf{b$_{1}$}=(\pi /d_{0},0)$ and $%
\textbf{b$_{2}$}=(0,\pi /d_{0})$. This two-dimensional spin-spin interaction obeys $%
C_{4v}$ point group symmetry of the Bravais lattice.

Analogous to the one-dimensional case, here we consider periodic boundary conditions along both the $x$ and $y$ directions. Applying the Fourier transformation to the four spins in a unit cell
\begin{eqnarray}
A_{n,m} &=&\frac{1}{\sqrt{N}}\sum_\textbf{k}e^{i(k_{x}n+k_{y}m)}A_{\textbf{k}},  \notag \\
B_{n,m} &=&\frac{1}{\sqrt{N}}\sum_\textbf{k}e^{i(k_{x}n+k_{y}m)}B_{\textbf{k}}, \notag \\
C_{n,m} &=&\frac{1}{\sqrt{N}}\sum_\textbf{k}e^{i(k_{x}n+k_{y}m)}C_{\textbf{k}}, \notag \\
D_{n,m} &=&\frac{1}{\sqrt{N}}\sum_\textbf{k}e^{i(k_{x}n+k_{y}m)}D_{\textbf{k}},
\end{eqnarray}%
where $\textbf{k}=(k_{x},k_{y})$ is the wavenumber in the first Brillouin zone. If we define the
unitary operator $\psi _{\textbf{k}}=\left(
\begin{array}{cccc}
A_{\textbf{k}} & B_{\textbf{k}} & C_{\textbf{k}} & D_{\textbf{k}}%
\end{array}%
\right) ^{T}$, the two-dimensional SSH Hamiltonian can be rewritten as%
\begin{equation}
H_{SSH}^{(2D)}=\sum_{\textbf{k}}\psi _{\textbf{k}}^{\dagger }H(\textbf{k})\psi _{\textbf{k}}.
\end{equation}%
Then we obtain $4\times 4$ matrix form of the Hamiltonian in the $\textbf{k}$-space
\begin{equation}
H(\textbf{k})=\hbar\left(
\begin{array}{cccc}
0 & f(k_{x}) & f(k_{y}) & 0 \\
f^{\ast }(k_{x}) & 0 & 0 & f(k_{y}) \\
f^{\ast }(k_{y}) & 0 & 0 & f(k_{x}) \\
0 & f^{\ast }(k_{y}) & f^{\ast }(k_{x}) & 0%
\end{array}%
\right).
\end{equation}%
$f(k_{x})=J_{1}+J_{2}e^{-ik_{x}}$ describes the spin-spin couplings in the $x$ direction, i.e., $A$ $%
\leftrightarrow $ $B$ and $C$ $\leftrightarrow $ $D$. While $%
f(k_{y})=J_{1}+J_{2}e^{-ik_{y}}$ represents the spin-spin couplings in the $y$ direction, i.e., $A$ $%
\leftrightarrow $ $C$ and $B$ $\leftrightarrow $ $D$.

\begin{figure}[tbp]
\includegraphics[width=8.6cm]{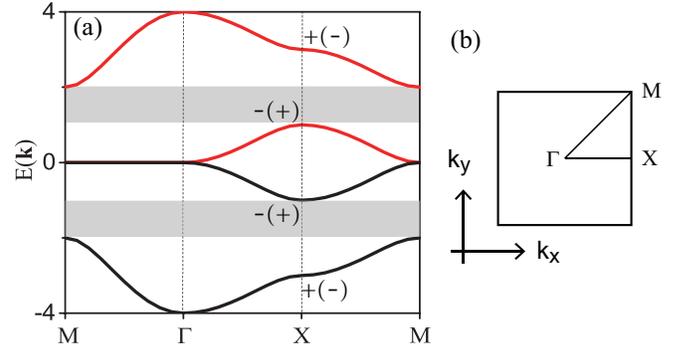}
\caption{(Color online) (a) Dispersion relation along the path $M\rightarrow \Gamma
\rightarrow X\rightarrow M$ in the first Brillouin zone. ``$\pm$" means the parity of the eigenstates under $\pi$ rotation. The gray area denotes the energy band gap. Note that the band gap will be vanished if $\delta=0$.
Here $\protect\delta =0.5$.(b) First Brillouin zone of the squire lattice.}
\label{fig_energy}
\end{figure}

We now study the dispersion relation of the 2D SSH model.
Solving the eigenvalue equation%
\begin{equation}
H(\textbf{k})\psi _{\textbf{k}}=E(\textbf{k})\psi _{\textbf{k}},
\end{equation}%
we obtain%
\begin{equation}
E(\textbf{k})=\epsilon _{x}\hbar \left\vert J_{1}+J_{2}e^{-ik_{x}}\right\vert +\epsilon
_{y}\hbar \left\vert J_{1}+J_{2}e^{-ik_{y}}\right\vert,
\end{equation}%
with $\epsilon _{x}=\epsilon _{y}=\pm 1$. In Figure.~$6$, we numerically calculate the band
structure of the 2D SSH model in momentum space. The energy spectrum
contains four bands: one around $E(\textbf{k})=4$, the symmetric one around $E(\textbf{k})=-4$,
and a pair of symmetric bands around $E(\textbf{k})=0$.

In addition, we simulate the dispersion relation in the first Brillouin
zone. As shown in Fig.~$7(a)$, the four energy bands can be grouped as: two
symmetric bands with opposite energies, and the middle pair  degenerate at $%
C_{4v}$ invariant points. There are two equal energy band gaps, with the width $%
\Delta E(\mathbf{k})=2\left\vert \delta \right\vert $. According to Eq.~$(41)
$, the system has the identical band structure if we swap the coupling
rates $J_{1}$ and $J_{2}$, which is the same as the one-dimensional case.
However, for the 2D SSH model, the eigenstates of the system
are qualitatively different. At the point of $X$, the wave functions possess
opposite parities for swapped $J_{1}$ and $J_{2}$. In Fig.~$7(a)$, we
label the opposite parities of the eigenstates as ``$+$" and ``$-$", respectively.

For the 1D SSH model, the energy band is closed when $\delta =0$. In this case, one can implement topological phase
transitions. Inspired by this result, we continue to explore the topological phase of the 2D SSH model. In order to exactly characterize the topological
nontrivial phase of the 2D SSH model, we introduce topological
invariant Chern number $\mathcal{C}=(C_{x},C_{y})$ \cite{rmp-82-1959}, with%
\begin{equation}
\mathcal{C}=\frac{1}{2\pi }\int_{BZ}dk_{x}dk_{y}\textbf{Tr}[A(\textbf{k})].
\end{equation}%
Here, $A(\textbf{k})=i\psi _{\textbf{k}}^{\dagger }\partial _{\textbf{k}}\psi _{\textbf{k}}$ is the non-Abelian Berry
connection, and the integration is performed over the first Brillouin zone (BZ).
Due to the $C_{4v}$ point group symmetry, we obtain $C_{x}=C_{y}$ in the
2D SSH system. In the case $\delta >0$, the Chern number $\mathcal{C}=(1/2
,1/2 )$, which implies that the system has a topological nontrivial phase.
However, in the case $\delta <0$, the Chern number $\mathcal{C}=(0,0)$, which correspondings
to the topological trivial case. Therefore, it is feasible to realize the topological phase
transition in the 2D SSH system by modulating the periodic driving.

\section{Quantum state transfer via the topological
edge states}

Long-range quantum state transfer is central to the study of time-evolving quantum systems \cite{rmp-70-1003,np-6-602,pra-79-042339}. As discussed above, applying a suitable periodic driving field to the SiV centers, the phononic band-gap mediated spin-spin interactions can be mapped to the SSH-type Hamiltonian. In the topological regime, we can obtain the obvious nontrivial edge states at the boundaries of the spin arrays. In this section, we take the 1D SSH as an example to show how quantum state  can be transferred with high fidelity via the topological edge states.

\subsection{Edge states}
The existence of edge states at the boundary is a distinguished feature for
topological insulator states \cite{prl-89-077002}. In the following, we will discuss how to
obtain the edge states in this spin-phononic system. We look for
the zero energy state in the 1D SSH model. Here we introduce the single-excited
state $|n_{\alpha }\rangle =\alpha _{n}^{\dagger }|ggg...\rangle $ $(\alpha
=A,B)$, which describes the spin at the $\alpha $ sites of the $n$th cell
excited to the state $|e\rangle $, while other spins stay in the ground state $%
|g\rangle $. In the single-excited state subspace, the Hamiltonian $H_{SSH}$
has the form
\begin{equation}
H_{SSH}=\overset{N}{\underset{n=1}{\sum }}\hbar(J_{1}|n_{A}\rangle \left\langle
n_{B}\right\vert +J_{2}|n_{B}\rangle \left\langle (n+1)_{A}\right\vert
+H.c.).
\end{equation}
Hence, we can get the the zero-energy eigenstates by sloving
\begin{equation}
H_{SSH}\overset{N}{\underset{n=1}{\sum }}(a_{n}|n_{A}\rangle
+b_{n}|n_{B}\rangle )=0,
\end{equation}%
where $a_{n}$ and $b_{n}$ are the amplitudes of occupying probability in the $%
n$th cell. There are $2N$ equations for the amplitudes $a_{n}$ and $b_{n}$,%
\begin{eqnarray}
J_{1}a_{n}+J_{2}a_{n+1} &=&0  \notag \\
J_{2}b_{n}+J_{1}b_{n+1} &=&0.
\end{eqnarray}%
\begin{figure}[tbp]
\includegraphics[width=8.6cm]{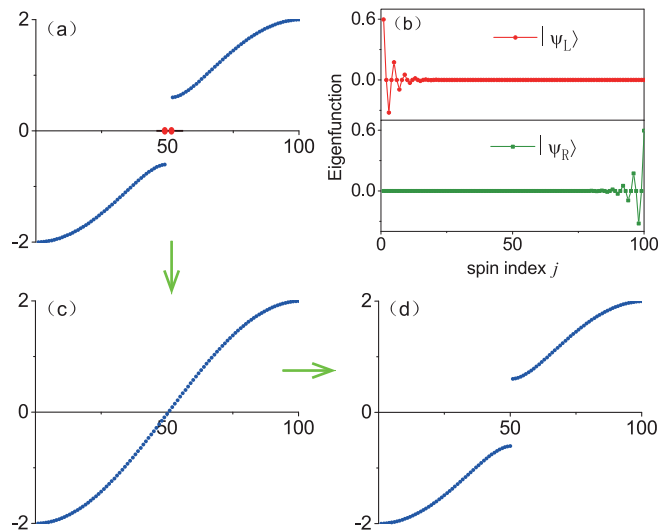}
\caption{(Color online) The single-excited energy spectrum of
Hamiltonian $H_{SSH}$ for different $\protect\delta$: (a) $\protect\delta %
=0.3$, (c) $\protect\delta =0$, (d) $\protect\delta =-0.3$. (b) Wavefunction
of the left and right edge states versus spin index $j$ with $\protect\delta %
=0.3$. The other parameters are $\mathcal{M}=1$ and $N=100$. }
\label{fig_energy}
\end{figure}
It should be noted that, for the boundaries, $b_{1}=a_{N}=0$. In the thermodynamic limit,
 $N\rightarrow \infty $, if we consider the case of $J_{1}<J_{2}$, we can
obtain the left and right zero-energy edge states as%
\begin{eqnarray}
|\psi _{L}\rangle &=&\overset{N}{\underset{n=1}{\sum }}a_{1}e^{-(n-1)/\xi
}|n_{A}\rangle ,  \notag \\
|\psi _{R}\rangle &=&\overset{N}{\underset{n=1}{\sum }}b_{N}e^{(n-N)/\xi
}|n_{B}\rangle ,
\end{eqnarray}%
where $\xi =1/\log (J_{1}/J_{2})$ is the localization length. When the ratio
$J_{2}/J_{1}$ becomes appreciably large, the wavefunction will almost be
confined at the first and last spins.

In order to verify the model, we numerically simulate the eigenvalues of the
system in Figs.~$8(a)$, $(c)$ and $(d)$. For $\delta >0$, i.e., $J_{1}<J_{2}$, there
exists a energy band gap and two zero-energy eigenvalues of the system. When
$\delta=0$, the gap is vanished. For $\delta <0$, the gap opens again but no
gapless modes appear. Correspondingly, we show the zero-energy edge states
in Fig.~$8(b)$. We see that the wavefunctions are localized exponentially in
the vicinity of the array edges, which is consistent with the theoretical
result.

As shown in Fig.~$8(b)$, we can also find that the left (right) edge states
only exist in the odd (even) spins, which is the consequence of chiral
symmetry \cite{lecture-2016}. In general, we say that a system with Hamiltonian $H$ has chiral
symmetry, if $\Gamma H\Gamma^{\dagger }=-H$, $\Gamma$ is the chiral symmetry operator. Here we define two orthogonal
projection operators,
\begin{equation}
\mathcal{P}_{A}=\frac{1}{2}(I+\Gamma),\mathcal{P}_{B}=\frac{1}{2}(I-\Gamma),
\end{equation}
where $I$ is the identity operator in the Hilbert space, $\mathcal{P}_{A}$ and $%
\mathcal{P}_{B}$ signify the projection to the spins at $A$ and $B$ sites,
respectively. Note that $\mathcal{P}_{A}+\mathcal{P}_{B}=1$ and $\mathcal{P}_{A}\mathcal{P}%
_{B}=0$. The Hamiltonian of the SSH model is bipartite: there are no
transitions between spins with the same label ($A$ or $B$), i.e., $\mathcal{P}%
_{A}H_{SSH}\mathcal{P}_{A}=\mathcal{P}_{B}H_{SSH}\mathcal{P}_{B}=0$. In fact, using the
projectors $\mathcal{P}_{A}$ and $\mathcal{P}_{B}$ is an alternative and equivalent
way of defining chiral symmetry. For the zero-energy eigenstates, we obtain%
\begin{equation}
H_{SSH}\mathcal{P}_{A/B}|\psi _{n}\rangle =H_{SSH}(|\psi _{n}\rangle \pm %
\Gamma|\psi _{n}\rangle )=0.
\end{equation}%
The projected zero energy states are eigenstates of the chiral symmetry operator $\Gamma$, and therefore are chiral symmetric partners of themselves. It is for this reason that the edge states are supported only by odd or even spins.

\subsection{Quantum state transfer}
In the following, we present the applications of this spin-phononic system and show that the topological edge states can be employed as a quantum channel between distance qubits. Since quantum information could be transferred directly between the boundary spins, the intermediate spins are virtually excited during the process, which ensures the robust quantum state transfer. Taking into account the coupling of the system with the environment in the Markovian approximation, the evolution of the system follows the master equation

\begin{equation}
\dot{\rho}=-\frac{i}{\hbar }[H_{SSH},\rho ]+\overset{2N}{\sum_{j=1}}\gamma
_{s}\mathcal{D}[\sigma _{j}^{z}]\rho ,
\end{equation}%
with $\sigma _{j}^{z}=|e\rangle _{j}\langle e|-|g\rangle _{j}\langle g|$, $%
\gamma _{s}$ the spin dephasing rate of the single SiV centers, and $\mathcal{D}[O]\rho
=O\rho O^{\dagger }-\frac{1}{2}\rho O^{\dagger }O-%
\frac{1}{2}O^{\dagger }O\rho $ for a given operator $O$.

\begin{figure}[tbp]
\includegraphics[width=8.6cm]{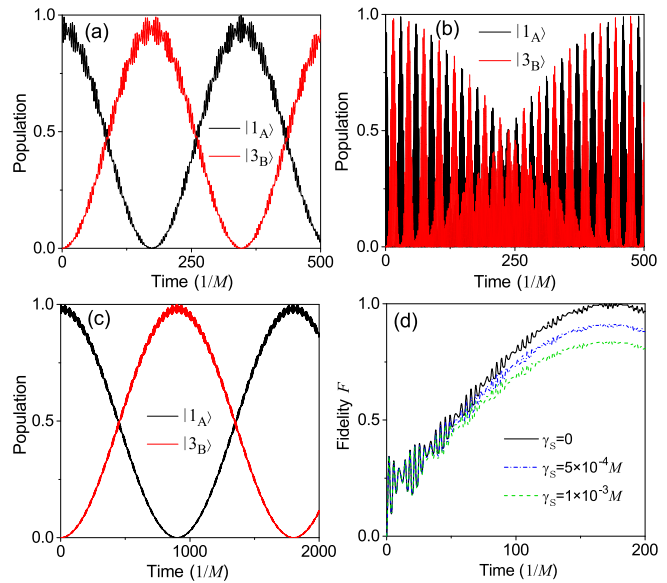}
\caption{(Color online) Time evolution of the population of the two edge states. (a) $%
\protect\delta =0.7$. (b) $\protect\delta =-0.7$. (c) $%
\protect\delta =0.82$. (d) Time evolution of the fidelity $F$ with different spin dephasing rate $%
\gamma _{s}$. The
other parameters are $\mathcal{M}=1$ and $N=3$. }
\label{fig_energy}
\end{figure}

To verify the transfer of the edge states and evaluate the performance of
this protocol, we numerically simulate the dynamics of the system by using
the QuTiP library, as shown in Fig. $9$. Here we take $6$ spins as an example,
i.e., $N=3$. As illustrated in Fig. $9(a)$, in the topological regime ($%
\delta >0$), we obtain the significant quantum state transfer between the two
end spins. However, for the non-topological condition, i,e., $\delta <0$, no
direct quantum state transfer can be seen, as shown in Fig. $9(b)$. In addition,
we simulate the effect of different values of the parameter $\delta $ on  quantum state transfer. Compared Fig. $9(a)$
with Fig. $9(c)$, we can see that the localization of the edge states is more obvious when $%
\delta $ takes a larger value, which is consistent with the theoretical
results. At the same time, for larger $\delta $, the time for accomplishing
quantum state transfer increases. It should be noted that the
time required for the system to realize the quantum state transfer should be shorter than the
coherence time of single SiV spins. Thus, when improving the value of $%
\delta $, we should consider the coherence time of the system as well. Note that
we neglect the spin dephasing in Fig. 9(a)-(c).

We now discuss the impact of spin dephasing of SiV centers on quantum information
transfer. Here we use fidelity to describe the performance of the state transfer, which is
defined as
\begin{equation}
\mathcal{F}=\left\{ Tr[(\sqrt{\rho (t_{f})}\rho (t_{i})\sqrt{\rho (t_{f})}%
)]^{1/2}\right\} ^{2}.
\end{equation}%
$\rho (t_{i})$ and $\rho (t_{f})$ denote the density operator for the
initial and final state of the transfer process, respectively \cite{rmap-9-273}. In Fig. 9(d), we present the
fidelity $\mathcal{F}$ as a function of time starting from the initial state $%
|1_{A}\rangle $. In the absence of spin dephasing, it is
shown that, the system evolves to the final state $|3_{B}\rangle $ with a
fidelity $\mathcal{F}\simeq 1$ (black solid line). This simulation result indicates
that quantum state transfer are indeed realized between the two end spins. Moreover, we simulate time evolution of the fidelity taking into consideration of
spin dephasing. As shown in Fig. 9(d), when setting the dephashing rate $%
\gamma _{s}=5\times 10^{-4}\mathcal{M}$, it is seen that the fidelity is about $%
0.91$. Furthermore, as the dephasing rate increases to $1\times 10^{-3}\mathcal{M}$%
, which is much closer to the realistic experimental conditions, the
fidelity of this scheme can still reach $\mathcal{F}=0.84$ (green dash line).
Therefore, our protocol can realize high fidelity quantum state transfer
with feasible experimental parameters.

\subsection{Approximate solutions}

\begin{figure}[tbp]
\includegraphics[width=8.6cm]{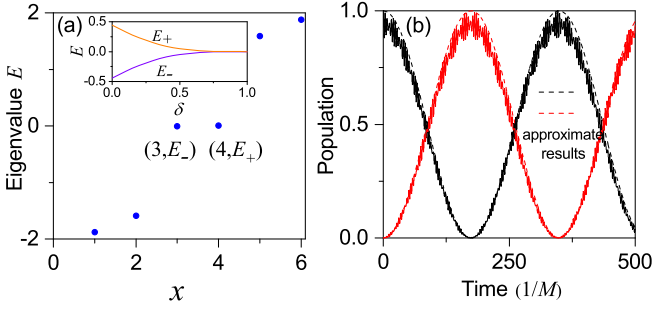}
\caption{(Color online) (a) $E_{+}$ and $%
E_{-}$ are the eigenvalues of the two edge states, and $%
\protect\delta =0.7$. The inset graph shows the eigenvalue $E$ as a function of $\delta$. (b) The comparison between the exact simulations and theoretical approximate results. Other parameters are the same as those in Fig. $9$.}
\label{fig_energy}
\end{figure}

In the following, we provide the comparison between our protocol and
theoretical approximate results. According to Eq.~$(46)$, if we assign
appreciably large values to the ratio $J_{2}/J_{1}$, the\ zero-energy edge
states will almost be confined at the first and end spins of the array. Here
we assume the two edge states $\left\vert e_{+}\right\rangle $ and $%
\left\vert e_{-}\right\rangle $, and the corresponding energies $E_{+}$ and $%
E_{-}$ are shown in Fig. 10(a). Then we can consider
\begin{eqnarray}
\left\vert 1_{A}\right\rangle  &=&\frac{1}{\sqrt{2}}(\left\vert
e_{+}\right\rangle +\left\vert e_{-}\right\rangle ),  \notag \\
\left\vert 3_{B}\right\rangle  &=&\frac{1}{\sqrt{2}}(\left\vert
e_{+}\right\rangle -\left\vert e_{-}\right\rangle ),
\end{eqnarray}%
If the initial condition is $\left\vert \psi (0)\right\rangle =\left\vert
1_{A}\right\rangle $, the time evolution of the quantum state follows
\begin{equation}
\left\vert \psi (t)\right\rangle =\frac{1}{\sqrt{2}}(e^{-iE_{+}t/\hbar
}\left\vert e_{+}\right\rangle +e^{-iE_{-}t/\hbar }\left\vert
e_{-}\right\rangle ).
\end{equation}%
Then we can obtain the mean population of the two ends of the array, i.e., $%
\left\langle 1_{A}\right\rangle =(1+\cos (\omega _{0}t))/2$ and $%
\left\langle 3_{B}\right\rangle =(1-\cos (\omega _{0}t))/2$ with $\omega
_{0}=(E_{+}-E_{-})/\hbar $. In Fig. $10(b)$, we plot the time evolution of the spin
populations, and find the excellent match between our protocol and
theoretical approximate results. It should be noted that, in non-topological regime, for
the same initial condition, the state will be a superposition of
more eigenstates and the particle will spread over the entire array.

\section{experimental consideration}

In this work, we consider a spin-phononic crystal system, where arrays of SiV centers are coupled by the quantized phonon modes of diamond phononic crystals. Based on state-of-the-art nanofabrication techniques, several experiments have demonstrated the generation of color center arrays through ion implantation \cite{NL-10-3168,prappl-7-064021}. And the fabrication of nanoscale mechanical structures with diamond crystals has been realized experimentally, as proposed in Refs. \cite{opt-3-1404,nat-462-78,NL-12-6048}. In general, the periodicity of a phononic crystal structure is characterised by periodic air holes etched on the crystal, which yields the tunable phononic bands. Since the excellent scalability of phononic crystal structures, this SSH model  is experimentally feasible when extending to the higher dimensional case.

For the diamond phononic crystal, the material properties are $E=1050$ GPa, $%
\nu =0.2$, and $\rho =3539$ kg/m$^{3}$. The lattice constant and cross
section of phononic crystal are $a=100$ nm and $A=100\times 20$ nm$^{2}$,
while the semi-major and -minor axis of the elliptical holes are $15$ nm and
$38$ nm, respectively. In this case, we get a phononic band edge frequency $%
\omega _{BE}/2\pi =44.933$ GHz, the ground state transition frequency of SiV
center is about $46$ GHz, which is exactly located in a phononic bandgap, as
shown in Fig. 2(a). The coupling between the SiV center and phononic crystal
mode $k$ is given by $g_{k}=\frac{d}{v_{l}}\sqrt{\frac{\hbar \omega _{BE}}{%
2\pi \rho aA}}\xi (\vec{r})$ \cite{NJP-17-043011,Natcomm-9-2012}, where $d/2\pi \sim 1$ PHz is the strain
sensitivity, and $v_{l}=1.71\times 10^{4}$ m/s is the speed of sound in diamond.
$\xi (\vec{r})$ is the dimensionless strain distribution at the position of
the SiV center $\vec{r}$, and here we assign $\xi (\vec{r})=1$ \cite{apl-87-043011}. Then we get the
SiV-phononic coupling rate $g_{k}/2\pi \simeq 100$ MHz. In the large
detuning regime, $g\sim 0.1g_{k}$, leading to the band gap engineered spin-phononic
coupling rate $g_{c}=g\sqrt{2\pi a/L_{c}}\simeq 2\pi \times 25$ MHz.

For the SiV color center in diamond, the two lower
sublevels $(|g\rangle ,|e\rangle )$ can be defined as a spin qubit and
coherently controlled by using microwave fields
 \cite{Natcomm-8-15579,prl-119-223602}. Moreover, in high-strain regime, the
magnetic dipole transition between the ground-state levels of SiV centers can be directly driven with microwaves, which is already experimentally
performed \cite{prb-100-165428}.
According to Eq.~$(11)$, the definition of $\delta $ is derived from the
periodic microwave driving. We can obtain different values of $\delta $ by
adjusting parameters $\eta $ and $\omega _{d}$ of the periodic driving
fields, which makes our model highly controllable and tunable.

At mK temperatures, the spin dephasing time of single SiV center is about $%
\gamma _{s}/2\pi =100$ Hz. As for phononic crystals, the mechanical quality
factor is $Q\sim 10^{7}$, which can be achieved and further improved by
using 2D phononic crystal shields \cite{prx-8-041027}. Thus we obtain the mechanical
dampling rate $\gamma _{m}/2\pi =4.5$ kHz. In this setup, the band gap
engineered spin-phononic coupling strength is $g_{c}/2\pi \simeq 25$ MHz,
which considerably exceeds both $\gamma _{s}$ and $\gamma _{m}$, resulting
in the strong strain interaction between the SiV centers and phonon crystal
modes. For the nearest neighbour spins with $d_{0}=a$, the phononic band-gap mediated
spin-spin interaction $\mathcal{M}=\frac{g_{c}^{2}}{2\Delta _{BE}}e^{-d_{0}/L_{c}}%
\simeq 2\pi \times 1.5$ MHz. For the quantum state transfer in Fig. 9(a), the period is $%
\mathcal{T}=160/\mathcal{M}\simeq 17$ $\mu$s, which is much shorter than the spin coherence time of SiV centers ($%
T_{2}^{\ast }\sim 10$ ms) \cite{prl-119-223602,Natcomm-8-15376}.

\section{conclusion}

In conclusion, we present a periodic driving protocol for realizing the SSH model in SiV-phononic crystal system. We study the band-gap engineered spin-phonon
coupling, and obtain the effective spin-spin interactions by adiabatically
eliminating the phonon modes. Then, in order to get the SSH model, we apply
a specific periodic driving to the SiV center spins. We discuss the
topological properties of the effective spin-spin system in momentum space,
and simulate the existence of the zero-energy topological edge states. In
addition, we study the long-range quantum
state transfer via topological edge states.

More importantly, compared with other
systems that simulate the SSH model, our scheme is more scalable and feasible in experimental
implementations. As an outlook, this scheme can be further extended to
higher dimensions. We can investigate the spin-phononic interaction in three-dimensional (3D) phononic crystals, and then study the corresponding topological properties. Moreover, since the inherent chiral symmetry of the SSH model, we can study the unidirectional quantum state transfer in this SiV-phononic crystal system. With the study of
phononic crystals, this proposal may be realized in near-future experiments,
and offers a realistic platform for the topological quantum computing and
quantum information processing.

\section*{Acknowledgments}

This work was supported by the NSFC under Grant No. 11774285, and the Fundamental Research
Funds for the Central Universities.

\end{document}